\documentclass[12pt]{article}
\usepackage{graphicx}
\usepackage{amsfonts}
\usepackage{amssymb,amsmath}
\usepackage{latexsym}
\usepackage{color}
\usepackage{cite}
\input{colordvi.tex}

\renewcommand{\thefootnote}{\#\arabic{footnote}}

\def\mathbi#1{\textbf{\em #1}}

\newcommand{\fnl}{f_{\rm NL}}
\newcommand{\gnl}{g_{\rm NL}}
\newcommand{\hnl}{h_{\rm NL}}
\newcommand{\inl}{i_{\rm NL}}
\newcommand{\jnl}{j_{\rm NL}}
\newcommand{\knl}{k_{\rm NL}}
\newcommand{\tnl}{\tau_{\rm NL}}

\newcommand{\calO}{{\cal O}}
\newcommand{\calP}{{\cal P}}

\begin{document}

\renewcommand{\thepage}{\arabic{page}}
\setcounter{page}{1}
\renewcommand{\thefootnote}{\#\arabic{footnote}}

\begin{titlepage}

\begin{center}

\hfill
APCTP Pre2013-011

\vskip .5in

{\Large \bf 
Higher order non-linear parameters with PLANCK
}

\vskip .45in

{\large
Jinn-Ouk~Gong$^{1,2}$ and Tomo~Takahashi$^3$
}

\vskip .45in

{\em
$^1$Asia Pacific Center for Theoretical Physics, Pohang 790-784, Korea \vspace{4 pt}\\
$^2$Department of Physics, Postech, Pohang 790-784, Korea \vspace{4 pt}\\
$^3$Department of Physics, Saga University, Saga 840-8502, Japan
}

\end{center}

\vskip .4in

\begin{abstract}

We investigate how higher order non-linear parameters affect  lower order ones 
through loop effects. 
We calculate the loop corrections up to two-loops and explicitly show 
that the tree contribution is stable against loop terms in most cases. 
We argue that, nevertheless, observational constraints on non-linear parameters such as $f_{\rm NL}$ and 
$\tau_{\rm NL}$ can also give a limit even for higher order ones due to the loop contribution. 
We discuss these issues for both single-source and multi-source cases.

\end{abstract}

\end{titlepage}

\setcounter{footnote}{0}

\section{Introduction}

Recent PLANCK data have well measured the quantities which characterize 
the nature of primordial density fluctuations with unprecedented accuracy \cite{Ade:2013rta,Ade:2013tta}.
Among them,  the non-linear parameter $f_{\rm NL}$, which parametrizes the amplitude of the bispectrum, 
has been of great interest since it was regarded as a critical test of inflation, and 
the PLANCK team obtained much severer constraints than pre-PLANCK observations.
For the so-called local type,  both $f_{\rm NL}$ and
$\tau_{\rm NL}$, which characterizes the size of the trispectrum for a certain configuration of wave numbers, 
 are now severely constrained as \cite{Ade:2013tta}
\begin{align}
 -8.9 < & f_{\rm NL}^{\rm (local)}  < 14.3 \, , \\
& \tau_{\rm NL} < 2800 \, ,
\end{align}
both at a 95\% confidence level, which is consistent with Gaussian density fluctuations\footnote{
The constraint on $g_{\rm NL}$ has  also been obtained, using WMAP9 data as 
$g_{\rm NL} = (-3.3 \pm 2.2)  \times 10^5$ 
at a 68\% confidence level \cite{Sekiguchi:2013hza}. For other works on the $\gnl$ constraints, 
see Ref.~\cite{gnl}.
}. 
Since the standard (single-field) inflation models predict almost Gaussian ones, 
they passed a critical test with the PLANCK results. Furthermore, other models of  
primordial fluctuations such as those utilizing a spectator field, like the curvaton model 
\cite{curvaton},
modulated reheating model \cite{modulated} and so on, which have been attracting attention 
due to their ability to produce large non-Gaussianity, are also still viable since they can also give 
$f_{\rm NL} \sim \mathcal{O}(1)$ in some parameter ranges.

However, one can also speculate a model in which the signature of non-Gaussianity comes not from 
the bispectrum but from higher order parameters such as $g_{\rm NL}$. In particular, in some models, 
large values of $g_{\rm NL}$ are possible even when $f_{\rm NL}$ satisfies the PLANCK constraint \cite{Suyama:2013nva}. 
In fact, when $g_{\rm NL}$ is large, it can contribute to the total value of $f_{\rm NL}$ through higher order correlation functions, or 
what is often called loop correction \cite{Lyth:1991ub}. In general, higher order non-linear parameters can also affect lower order ones 
through the loop corrections, which  may enable us to probe higher order non-linear parameters by using observational 
constraints on lower order counterparts such as $f_{\rm NL}$ and $\tau_{\rm NL}$. Since PLANCK put a stringent constraint 
on $f_{\rm NL}$ and even on $\tau_{\rm NL}$, it would be interesting to see to what extent one can 
probe higher order non-linear parameters using the PLANCK constraints\footnote{
See also Ref.~\cite{Byrnes:2013qjy} for the implication of the PLANCK $\fnl$ constraint on the trispectrum parameter.
}, which is the main topic in this article.

In the next section, we summarize the formalism and set our notations. We also give the explicit expressions of $f_{\rm NL}, \tau_{\rm NL}$, 
and $g_{\rm NL}$, including the loop correction up to two-loops.
After obtaining the  expressions for these non-linear parameters, in Section \ref{sec:probe}, 
we investigate the constraints for higher order non-linear parameters using the PLANCK constraints on $f_{\rm NL}$ and $\tau_{\rm NL}$.
The final section is devoted to conclusion of this article and discussion.

\section{Formalism}
\label{sec:formalism}

In this section, we give a formalism and the expressions for non-linear parameters 
including the loop contributions. We focus on the local-type model,
in which the curvature perturbation on the uniform density slice $\zeta$ may be expanded as
\begin{align}\label{zetaexpansion}
\zeta & 
= 
\zeta_g + \frac{3}{5}\fnl\zeta_g^2 + \left( \frac{3}{5} \right)^2\gnl\zeta_g^3 + \left( \frac{3}{5} \right)^3\hnl\zeta_g^4 
+  \left( \frac{3}{5} \right)^4\inl\zeta_g^5 +  \left( \frac{3}{5} \right)^5\jnl\zeta_g^6 
+  \left( \frac{3}{5} \right)^6\knl\zeta_g^7 + 
\cdots 
\nonumber\\
& \equiv \zeta^{(1)} + \zeta^{(2)} + \zeta^{(3)} + \zeta^{(4)} + \zeta^{(5)} +  \zeta^{(6)} + \zeta^{(7)} + \cdots \, ,
\end{align}
where we have defined explicitly the ``bare'' non-linear parameters as above and
included up to the 7th order: this is required to calculate 
the corrections up to two-loop order for the trispectrum.
For this perturbative expansion to be valid, we may naturally require $\zeta^{(n)} \gtrsim \zeta^{(n+1)}$. 
Thus, with $\zeta_g \sim 2 \times 10^{-5}$, we obtain a generous bound for each non-linear parameter smaller than $10^4$-$10^5$. 
Notice that one may relax this assumption in such a way that all the non-linear terms are 
of the same order of magnitude, $\zeta^{(2)} \sim \zeta^{(3)} \sim \zeta^{(4)} \cdots$,
\`a la general slow-roll approximation~\cite{gsr}. In this case,
the higher order non-linear parameters may be far larger than $10^5$. 
However, more generally, for such a hierarchy we need very elaborated models
and we do not consider this possibility here.

The power spectrum $P_\zeta (k)$, bispectrum $B_\zeta (k_1, k_2, k_3)$, and 
trispectrum $T_\zeta (k_1, k_2, k_3, k_4)$ are defined as
\begin{align}
\left\langle \zeta(\mathbi{k}_1)\zeta(\mathbi{k}_2) \right\rangle 
& \equiv (2\pi)^3\delta^{(3)}(\mathbi{k}_1+\mathbi{k}_2) P_\zeta(k_1) \, ,  \\
\left\langle \zeta(\mathbi{k}_1)\zeta(\mathbi{k}_2) \zeta(\mathbi{k}_3) \right\rangle 
& \equiv (2\pi)^3\delta^{(3)}(\mathbi{k}_1+\mathbi{k}_2 +\mathbi{k}_3) B_\zeta(k_1,k_2,k_3) \, ,  \\
\left\langle \zeta(\mathbi{k}_1)\zeta(\mathbi{k}_2) \zeta(\mathbi{k}_3) \zeta(\mathbi{k}_4) \right\rangle 
& \equiv (2\pi)^3\delta^{(3)}(\mathbi{k}_1+\mathbi{k}_2 +\mathbi{k}_3 +\mathbi{k}_4) \, 
T_\zeta(k_1,k_2,k_3,k_4) \, .
\end{align}
Below we explicitly give the expressions for these spectra including the corrections 
up to two-loops, and we will see the effects of the higher order non-linear parameters through these loops.

\subsection{Power spectrum}

Introducing the notation 
\begin{equation}
\left\langle \zeta^{(i)}(\mathbi{k}) \zeta^{(j)}(\mathbi{q}) \right\rangle + \text{possible perm}
= (2\pi)^3 \delta^{(3)}(\mathbi{k}+\mathbi{q})  P_\zeta^{(ij)}(k) \, , 
\end{equation}
which represents the contribution from the correlation function of the $i$th- and $j$th order 
terms in \eqref{zetaexpansion}, the power spectrum up to two-loop corrections is given by
\begin{equation}
P_\zeta (k) = 
P_\zeta^{(11)}  
+ P_\zeta^{(13)}+ P_\zeta^{(22)}
+ P_\zeta^{(15)} + P_\zeta^{(24)} + P_\zeta^{(33)} \, ,
\end{equation}
where $P_\zeta^{(11)}$ gives the tree term, 
$P_\zeta^{(22)}$ and $P_\zeta^{(13)}$ constitute one-loop~\cite{Gong:2010yk}
and $P_\zeta^{(15)}, P_\zeta^{(24)}$, and $P_\zeta^{(33)}$ correspond to two-loop corrections. 
Explicit expressions for these terms are given in Appendix~\ref{app:power}.
Writing the loop corrections in terms of the non-linear parameter and the tree power spectrum explicitly,
we find
\begin{align}
P_\zeta(k) 
&=
P_\zeta^{(11)}(k) \left\{ 1 + \left(\frac35 \right)^2 \left(4\fnl^2 + 6\gnl \right) \calP_\zeta(L^{-1})\log(kL)
\right. \notag \\
&
\left. \hspace{2cm}
+ \left(\frac35 \right)^4 \left( 48 \fnl \hnl + 45 \gnl^2 + 30\inl \right)
\left[ \calP_\zeta(L^{-1})\log(kL) \right]^2
 \right\} \, ,
\end{align}
where $L$ is a large fictitious box size in which the Fourier modes of the curvature perturbation are taken and we have assumed $\calP_\zeta \equiv k^3P_\zeta^{(11)}/\left(2\pi^2\right) \sim 2.5 \times 10^{-9}$ remains nearly scale invariant over this box. Over the observable scales, the logarithm gives $\calO(1)$.

\subsection{Bispectrum}

The bispectrum up to two-loop corrections is given by
\begin{align}
\label{eq:bi-1}
B_\zeta(k_1,k_2,k_3) 
= &
B_\zeta^{(112)} \notag \\
&
+ B_\zeta^{(114)} + B_\zeta^{(123)}  + B_\zeta^{(222)}  \notag \\
&
+ B_\zeta^{(116)} + B_\zeta^{(125)} + B_\zeta^{(134)}  + B_\zeta^{(224)} + B_\zeta^{(233)} 
\, ,
\end{align}
where we have defined the notation, and likewise for the case of the power spectrum, 
\begin{equation}
\left\langle \zeta^{(i)}(\mathbi{k}_1) \zeta^{(j)}(\mathbi{k}_2)  \zeta^{(k)}(\mathbi{k}_3)  \right\rangle + \text{possible perms}
= (2\pi)^3 \delta^{(3)}(\mathbi{k}_1+\mathbi{k}_2 +\mathbi{k}_3) B_\zeta^{(ijk)} \, .
\end{equation}
In (\ref{eq:bi-1}), the first, second and third lines on the right hand side correspond to the tree, 
one-loop  and  two-loop contributions, respectively. Explicit expressions for these
terms are given in Appendix~\ref{app:bi}.
Putting everything together, $\fnl$ in the squeezed limit including up to two-loop correction is given by\footnote{
The expression of $\fnl$ up to one-loop has been obtained in Ref.~\cite{Tasinato:2012js}.
}
\begin{align}
\label{eq:fnl-1}
f_{\rm NL}^{\rm (tot)} 
\equiv &
\frac{5}{12} \lim_{k_3\to0} \frac{B_\zeta(k_1,k_2,k_3)}{P_\zeta(k_1)P_\zeta(k_3)}  
\notag \\
=&    
  f_{\rm NL}
  + \left( \frac35 \right)^3
   \left(-\frac{20}{3}  f_{\rm NL}^3+10  f_{\rm NL} g_{\rm NL}
   +10 h_{\rm NL} \right) \left[ \calP_\zeta(L^{-1})\log(kL) \right]  \notag \\
     &  +
     \left( \frac35 \right)^5 
 \left(\frac{80}{3}  f_{\rm NL}^5 - 80  f_{\rm NL}^3 g_{\rm NL}
  - 40 f_{\rm NL}^2 h_{\rm NL}+ 45  f_{\rm NL} g_{\rm NL}^2 
  +150 f_{\rm NL}
   i_{\rm NL}
\right.   \notag \\
& 
\left. \hspace{2cm}
   +180 g_{\rm NL} h_{\rm NL}+ 75 j_{\rm NL}\right)
    \left[ \calP_\zeta(L^{-1})\log(kL) \right]^2 \, .
\end{align}
Notice that $f_{\rm NL}$ on the right hand side of the above expression is the ``bare" $f_{\rm NL}$ appearing in (\ref{zetaexpansion})
which should not be confused with $\fnl^{\rm (tot)}$, that includes loop corrections.

\subsection{Trispectrum}

For the trispectrum, including up to two-loop corrections, we obtain 
\begin{align}
\label{eq:tri-1}
& T_\zeta(k_1,k_2,k_3,k_4) 
\notag \\
= &
T_\zeta^{(1113)}   + T_\zeta^{(1122)} \notag \\
&
+ T_\zeta^{(1115)}  + T_\zeta^{(1124)}  + T_\zeta^{(1133)}   + T_\zeta^{(1223)} + T_\zeta^{(2222)} 
\notag \\
&
+ T_\zeta^{(1117)} + T_\zeta^{(1126)} + T_\zeta^{(1135)}  + T_\zeta^{(1144)} + T_\zeta^{(1225)} 
+ T_\zeta^{(1234)} + T_\zeta^{(1333)} + T_\zeta^{(2224)} + T_\zeta^{(2233)}  \, ,
\notag \\
\end{align}
where the first, second and third lines on the right hand side correspond to  the tree, 
one-loop  and  two-loop contributions, respectively. 
Here we have again introduced a notation $T_\zeta^{(ijkl)}$ as 
\begin{equation}
\left\langle 
\zeta^{(i)}(\mathbi{k}_1) \zeta^{(j)}(\mathbi{k}_2)  \zeta^{(k)}(\mathbi{k}_3) \zeta^{(l)}(\mathbi{k}_4)
\right\rangle + \text{possible perms}
= (2\pi)^3 \delta^{(3)}(\mathbi{k}_1+\mathbi{k}_2 +\mathbi{k}_3 + \mathbi{k}_4) T_\zeta^{(ijkl)} \, .
\end{equation}
Explicit expressions for these terms are given in Appendix~\ref{app:tri}.

There are two non-linear parameters corresponding to 
different dependences on $k_{ij} \equiv |\mathbi{k}_i+\mathbi{k}_j|$. 
Collecting $k_{ij}$ dependence in the trispectrum, 
$\tnl$ is defined in the collapsed limit, 
including two-loop corrections, as\footnote{
The expression of $\tnl$ up to one-loop has been obtained in Ref.~\cite{Tasinato:2012js}.
} 
\begin{align}
\label{eq:tnl-1}
\tnl^\text{(tot)} 
=&  \frac{1}{4} \lim_{k_{12}\to0} \frac{T_\zeta \text{ with $k_{ij}$ dependence}}{P_\zeta(k_1)P_\zeta(k_3)P_\zeta(k_{12})} 
\nonumber \\
=&
 \left( \frac56 \fnl \right)^2  
+
\left( \frac35 \right)^4 \left( 
-32 \fnl^4 +48 \fnl^2 \gnl + 36 \gnl^2 + 48 \fnl \hnl
\right)   \left[ \calP_\zeta(L^{-1})\log(kL) \right]  \notag \\
&
+
\left( \frac35 \right)^6 \left( 
192 \fnl^6 - 576 \fnl^4 \gnl + 1404  \fnl^2 \gnl^2 
-  384 \fnl^3 \hnl + 3456 \fnl \gnl \hnl 
\right. \notag  \\
&
\left. \hspace{1.8cm}
+ 720 \hnl^2  +  720 \fnl^2 \inl + 720 \gnl \inl + 360 \fnl \jnl
\right)  \left[ \calP_\zeta(L^{-1})\log(kL) \right]^2
\end{align}
On the other hand, another trispectrum parameter  $\gnl$ is defined for those without $k_{ij}$ 
dependence in the doubly squeezed limit as
\begin{align}
\label{eq:gnl-1}
\gnl^\text{(tot)} 
= &
\frac{25}{108} \lim_{k_1,k_2\to0} \frac{T_\zeta \text{ without $k_{ij}$ dependence}}{P_\zeta(k_1)P_\zeta(k_2)P_\zeta(k_{3})} 
\nonumber\\
= &
\gnl
+\left( \frac35 \right)^2 \left( 
-9 \gnl^2 + 24 \fnl \hnl + 10 \inl
\right)  \left[ \calP_\zeta(L^{-1})\log(kL) \right] \notag \\
&
+\left( \frac35 \right)^4 \left(  
-48 \fnl^4 \gnl + 180 \fnl^2 \gnl^2 + 54 \gnl^3 
- 192 \fnl^3 \hnl + 288 \fnl \gnl \hnl 
\right. \notag  \\
&
\left. \hspace{1.8cm}
+ 144 \hnl^2 
+  240 \fnl^2 \inl + 225 \gnl \inl + 360 \fnl \jnl + 105 \knl
\right)  \left[ \calP_\zeta(L^{-1})\log(kL) \right]^2. \notag \\
\end{align}
With these machineries, in the next section, we discuss implications of 
constraints on $\fnl^{\rm (tot)}$ and $\tnl^{\rm (tot)}$ for higher order non-linear parameters.

\section{Probing higher order non-linear parameters}
\label{sec:probe}

\subsection{Single source case}

\begin{figure}[htbp]
  \begin{center}
    \resizebox{160mm}{!}{
     \includegraphics{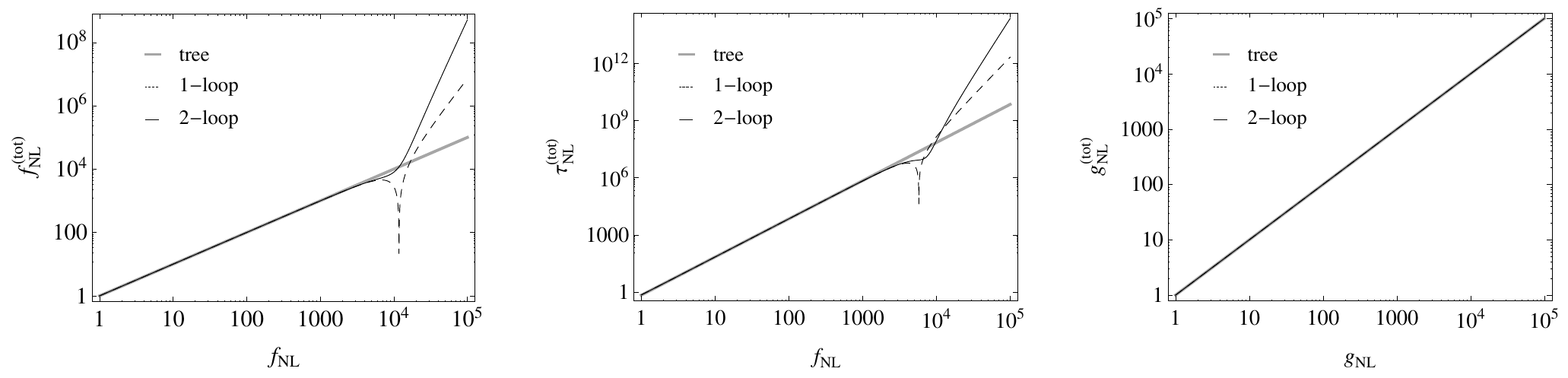}
}
  \end{center}
  \caption{Plot of  $f_{\rm NL}^{\rm (tot)}$ (left), $\tnl^{\rm (tot)}$ (middle) and $\gnl^{\rm (tot)}$ (right) as a function of 
  bare non-linear parameters. Here absolute values are plotted. In all panels, we assume $h_{\rm NL}=10^3$ and set all higher non-linear parameters to be zero.
  In the left and middle panels,  we set $g_{\rm NL}=10^3$, and in the right panel, $\fnl = 10^3$ is assumed.
     However, as long as we set all non-linear parameters less than $10^5$, we obtain more or less the same results. 
    Notice that at large $f_{\rm NL}$, the sign of $f_{\rm NL}^{\rm (tot)}$ and $\tnl^{\rm (tot)} $ changes from positive to negative
    and a cancellation 
 occurs to give $f_{\rm NL}^{\rm (tot)}$ and $\tnl^{\rm (tot)} $ vanishing at one-loop level,  however, when two-loop corrections are included, 
 such a cancellation does not occur.}
  \label{fig:fNL-1-2-loop}
\end{figure}

First we consider $\fnl^\text{(tot)}$ as a function of bare $\fnl$ with the other non-linear
parameters being fixed (see later), which is shown in the left panel of Fig.~\ref{fig:fNL-1-2-loop}. 
Up to $\fnl \sim 10^4$, the tree result, i.e. $\fnl^\text{(tot)}=\fnl$, holds very accurately. 
However, for $\fnl \sim 10^4$ if we truncate at one-loop level, the tree result $\fnl$ 
is canceled by the contribution from the $\fnl^3$ term since the sign is opposite. 
Meanwhile if we further include the two-loop contributions, such a cancellation does
not occur and one may conclude that the truncation at a certain order may lead to 
misevaluation of non-linear parameters. But at this large value of the bare $\fnl$ the 
perturbativity of the expansion (\ref{zetaexpansion}) is broken, 
i.e. the hierarchy $\zeta^{(1)} \gtrsim \zeta^{(2)}$ does not hold,
 and we can no longer 
be sure that the loop corrections are under control. In other words, as long as the bare
$\fnl$ is not too large to harm the perturbative expansion (\ref{zetaexpansion}), 
the tree term is absolutely dominating over loop corrections. We can draw similar results
for $\tnl$ and $\gnl$ shown in the middle and right panels of Fig.~\ref{fig:fNL-1-2-loop}.

\begin{figure}[htbp]
    \resizebox{160mm}{!}{
     \includegraphics{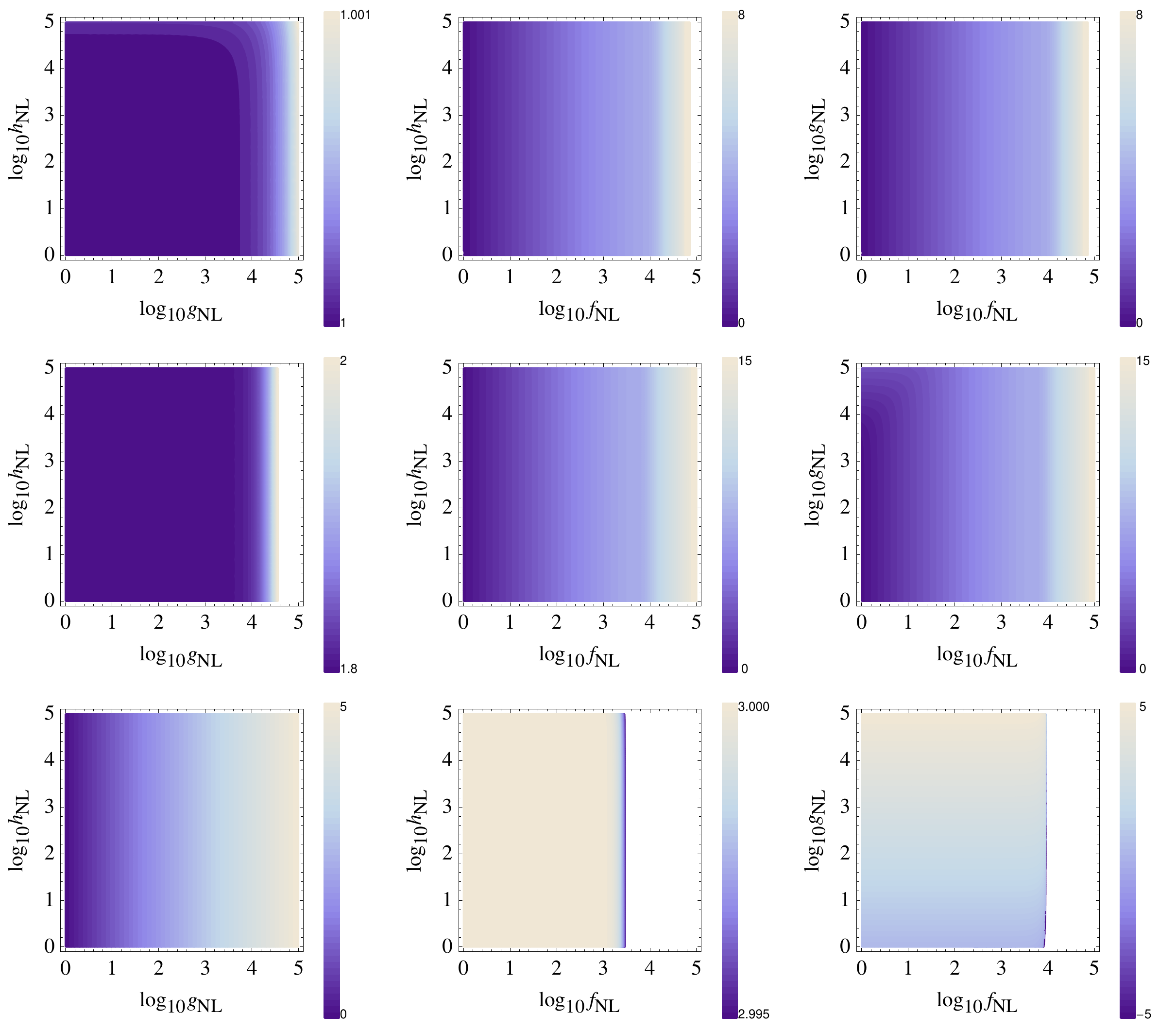}
}
  \caption{Contour plots of $\fnl^{\rm (tot)}$ (top), $\tnl^{\rm (tot)}$ (middle) and $\gnl^{\rm (tot)}$ (bottom).
  We set $\fnl=10$, $\gnl=10^3$, and $\hnl = 10^3$ for the left, middle, and right columns, respectively. 
  In all panels, other non-linear parameters are set to be zero. 
  The labels next to the color bar at the right of each panel show the values in units of $\log_{10}$.
  We note here that  $\fnl^{\rm (tot)}$ and $\tnl^{\rm (tot)}$
  are more enhanced due to two-loop contribution compared to those from the tree one 
  for the values of $\fnl$ larger than $10^4$ as also seen from Fig.~\ref{fig:fNL-1-2-loop}. 
  Thus, we show contours up to $\fnl^{\rm (tot)} =10^8$ and $\tnl^{\rm (tot)}$ in some panels.
  }
  \label{fig:fNL_loop}
\end{figure}

\begin{figure}[htbp]
  \begin{center}
    \resizebox{100mm}{!}{
      \includegraphics{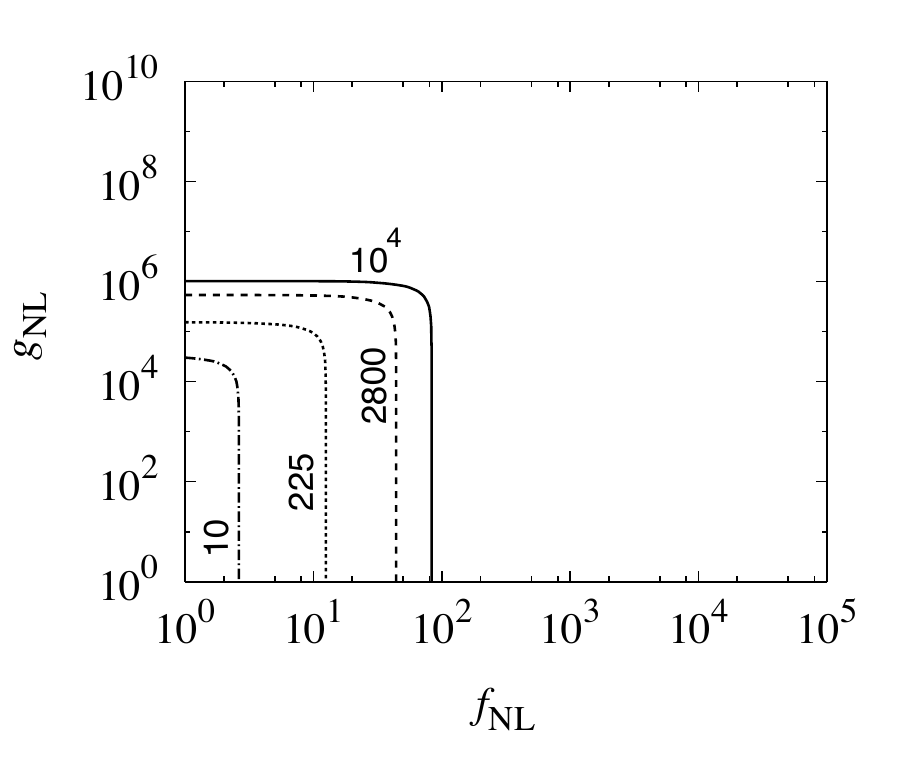}
}
  \end{center}
  \caption{Contours of $\tau_{\rm NL}^{\rm (tot)}$  
  in the $f_{\rm NL}$--$g_{\rm NL}$ plane. This plot is the same as the rightmost one in the middle row of Fig.~\ref{fig:fNL_loop}, but 
  with $h_{\rm NL}=0$. Here we give a few noticeable contours of $\tau_{\rm NL}^{\rm (tot)}$ explicitly.
  Especially, the PLANCK bound ($\tau_{\rm NL}^{\rm (tot)}<2800$) and (expected) EPIC sensitivity ($\tau_{\rm NL}^{\rm (tot)}<255$)
  are shown. }
  \label{fig:tNL_loop}
\end{figure}

Next, we show the contours of $\fnl^\text{(tot)}$, $\tnl^\text{(tot)}$, and $\gnl^\text{(tot)}$ 
as functions of bare $\fnl$, $\gnl$, and $\hnl$ in Fig~\ref{fig:fNL_loop}. 
The values of other non-linear parameters assumed in the figure are shown in the caption.
 As seen from the plots, the loop corrected values of these non-linear
parameters are almost determined by the corresponding bare ones, as long as higher order
non-linear parameters are not too large to respect the hierarchy  of the non-linear
expansion (\ref{zetaexpansion}). 
We note here that in principle, higher order non-linear parameters can be 
constrained from observations of $\fnl^\text{(tot)}$, although their constraints are very weak.
Nevertheless, it is interesting to see that,
adopting $\tnl < 2800$ (95 \% C.L. from PLANCK data),  we obtain $\gnl < 5 \times 10^5$, 
which is almost the same as the actual constraint from WMAP9, $\gnl = (-3.3 \pm 2.2)  \times 
10^5~ {\rm (68\%~C.L.)}$ \cite{Sekiguchi:2013hza}. In the future, more severe constraints would be obtained. For example, 
projected constraints on non-linear parameters  from EPIC have been investigated in Ref.~\cite{Smidt:2010ra}, 
where 1$\sigma$ constraints on $\tnl$ and $\gnl$ are given as
$\Delta \tnl =225$ and $\Delta \gnl = 6.0 \times 10^4$. From  Fig.~\ref{fig:tNL_loop}, we can again see that 
a constraint on $\tnl$ at this level also gives a similar limit on $\gnl$ through loop corrections.

\subsection{Mixed source case I}

\begin{figure}[htbp]
  \begin{center}
    \resizebox{165mm}{!}{
     \includegraphics{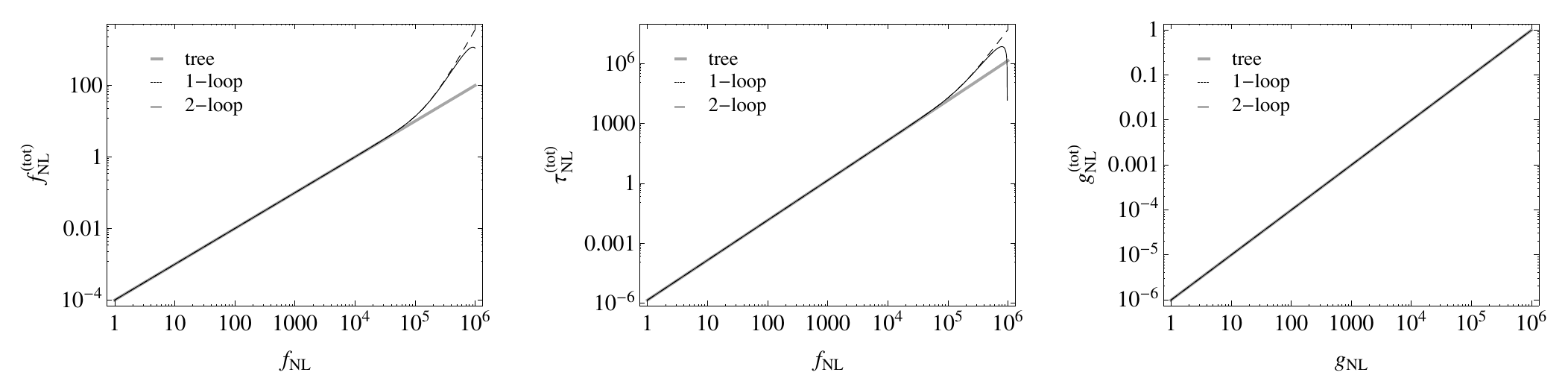}
}
  \end{center}
  \caption{Plot of  $f_{\rm NL}^{\rm (tot)}$ (left), $\tnl^{\rm (tot)}$ (middle) and $\gnl^{\rm (tot)}$ (right)  as a function of 
  bare non-linear parameters. For other non-linear parameters, we assume the same as in Fig.~\ref{fig:fNL-1-2-loop}. 
  Here we assume $R=0.01$.  However, as long as we set all non-linear parameters to be less than $10^5$, we obtain more or less the same results. 
 }
  \label{fig:2comp_bareNLvsloopNL}
\end{figure}

\begin{figure}[htbp]
  \begin{center}
    \resizebox{140mm}{!}{
     \includegraphics{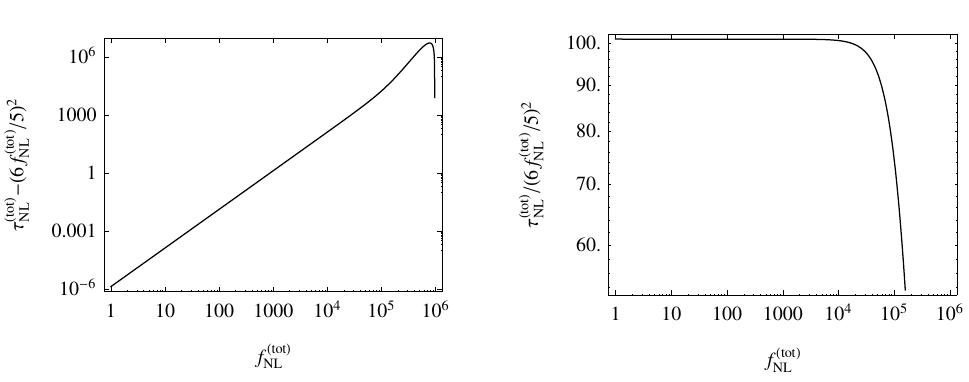}
}
  \end{center}
  \caption{Plot of $\tnl^{\rm (tot)} - (6 \fnl^{\rm (tot)}/5)^2$ (left) and $\tnl^{\rm (tot)} /(6 \fnl^{\rm (tot)}/5)^2$ (right)  for a multi-field model. Here we set non-linear parameters as the same as
  in Fig.~\ref{fig:2comp_bareNLvsloopNL}.   
  The Suyama-Yamaguchi inequality holds as long as the perturbative expansion is valid.}
  \label{fig:2compSYineq}
\end{figure}

Even when we consider a spectator field model such as the curvaton, modulated reheating and so on, 
an inflaton field should exist to drive an inflationary expansion, and it also should have fluctuations. 
When the fluctuations from the inflaton can be neglected, a spectator field model can be 
regarded as the single-field case, which was discussed in the previous section: we can apply
(\ref{eq:fnl-1}), (\ref{eq:tnl-1}) and (\ref{eq:gnl-1}) for the non-linear parameters $\fnl^\text{(tot)}$, 
$\tnl^\text{(tot)}$, and $\gnl^\text{(tot)}$. 
However, to discuss more general cases, we should consider fluctuations from both the inflaton 
and a spectator field. This kind of model has been called a ``mixed models"
and investigated for the curvaton  \cite{mixedcurvaton}, the modulated reheating model \cite{Ichikawa:2008ne} and for 
a general case in the light of recent PLANCK data \cite{Enqvist:2013paa}.
In this case, the curvature perturbation can be given by 
\begin{align}
\label{zetaexpansion-spectator}
\zeta & 
= 
\zeta_\phi  + \zeta_\sigma + \frac{3}{5}\fnl\zeta_\sigma^2 + \left( \frac{3}{5} \right)^2\gnl\zeta_\sigma^3 
+ \left( \frac{3}{5} \right)^3\hnl\zeta_\sigma^4 
+ \cdots \, ,
\end{align}
where $\zeta_\phi$ and $\zeta_\sigma$ are the Gaussian parts of the curvature perturbation 
from the inflaton $\phi$ and the spectator field $\sigma$, respectively.  
For the inflaton part, we have neglected higher order contributions since they are slow-roll suppressed for a standard inflation model.
Then, 
the power spectrum is given by 
\begin{align}
\label{eq:power_mixed}
P_\zeta(k) 
=& 
P_\phi (k) \left\{ 1 + R +  \left(\frac35 \right)^2 \left(4\fnl^2 + 6\gnl \right) R^2 \calP_\phi(L^{-1})\log(kL)
\right. \notag \\
&
\left. \hspace{1.3cm}
+ \left(\frac35 \right)^4 \left( 48 \fnl \hnl + 45 \gnl^2 + 30\inl \right) 
R \left[ R \calP_\phi (L^{-1})\log(kL) \right]^2
 \right\} \, .
\end{align}
Here we have introduced a notation representing the ratio of the power spectra 
generated from the inflaton $\phi$ and the spectator field $\sigma$ at some reference scale as 
\begin{equation}
\label{eq:R}
R \equiv \frac{P_\sigma}{P_\phi} \, , 
\end{equation}
where $P_\phi$ and $P_\sigma$ are defined as 
\begin{align}
\left\langle \zeta_\phi (\mathbi{k}) \zeta_\phi (\mathbi{q}) \right\rangle
& =   (2\pi)^3 \delta^{(3)} ( \mathbi{k} + \mathbi{q} ) P_\phi (k) \, ,
\\
\left\langle \zeta_\sigma (\mathbi{k}) \zeta_\sigma (\mathbi{q}) \right\rangle
& =   (2\pi)^3 \delta^{(3)} ( \mathbi{k} + \mathbi{q} ) P_\sigma (k) \, .
\end{align}
Non-linear parameters are then found to be
\begin{align}
\label{eq:fnl-mixed}
f_{\rm NL}^{\rm (tot)} 
=&    
\left( \frac{R}{1+R} \right)^2
\left\{  f_{\rm NL}
  + \left( \frac35 \right)^3
  \left[ \frac{20}{3}  f_{\rm NL}^3+30  f_{\rm NL} g_{\rm NL}
   +10 h_{\rm NL} \right. \right. \notag \\
   &
   \left. \left.
   \qquad\qquad\qquad\qquad
+ R   \left(-\frac{20}{3}  f_{\rm NL}^3+10  f_{\rm NL} g_{\rm NL}
   +10 h_{\rm NL} \right) \right] \frac{R}{1+R} \calP_\phi(L^{-1})\log(kL) 
   \right.
   \notag \\
     &
     \left. 
      +
     \left( \frac35 \right)^5 \left[
     25 \left( 
     8 \fnl^2 \hnl + 15 \fnl \gnl^2 + 12 \gnl \hnl + 10 \fnl \inl + 3\jnl
     \right) 
     \right. \right.  \notag\\
     &
     \left. \left.
     + 10R \left( 
     - \frac{16}{3}  \fnl^5   - 32 \fnl^3 \gnl +  24 \fnl \gnl^2  + 16 \fnl^2 \hnl + 48 \gnl \hnl + 40 \fnl \inl + 15 \jnl
     \right) 
       \right. \right.  \notag\\
     &
     \left. \left.
+  R^2\left(\frac{80}{3}  \fnl^5 - 80  \fnl^3 \gnl  + 45 \fnl^2 \gnl -40 \fnl^2 \hnl + 180 \gnl \hnl 
  +150 \fnl \inl + 75 \jnl \right) \right]
\right.    \notag \\
& 
\left.   
\qquad\qquad\qquad\qquad\qquad\qquad\qquad\qquad\qquad
      \times \left[ \frac{R}{1+R}  \calP_\phi(L^{-1})\log(kL) \right]^2
    \right\} \, ,
\end{align}
\begin{align}
\label{eq:tnl-mixed}
\tnl^\text{(tot)} 
=&
\left( \frac{R}{1+R} \right)^3 
\left\{
 \left( \frac56 \fnl \right)^2  
+
\left( \frac35 \right)^4 \left[
16 \fnl^4 +120 \fnl^2 \gnl + 36 \gnl^2 + 48 \fnl \hnl 
\right.\right. \notag \\
&
\left.\left.
+R \left( -32 \fnl^4 +48 \fnl^2 \gnl + 36 \gnl^2 + 48 \fnl \hnl   \right)
\right]    \frac{R}{1+R}  \calP_\phi (L^{-1})\log(kL) 
\right.
\notag \\
&
\left.
+
\left( \frac35 \right)^6 \left[
3672 \fnl^2  \gnl^2 + 648  \gnl^3 + 768  \fnl^2 \hnl + 4320 \fnl \gnl \hnl  + 720 \hnl^2 
\right.\right. \notag \\
&
\left.\left. 
\qquad
+ 1080 \fnl^2 \inl + 720 \gnl \inl + 360 \fnl \jnl
\right.\right. \notag \\
&
\left.\left. 
+ R \left(- 192 \fnl^6 - 1728 \fnl^4 \gnl + 4212  \fnl^2 \gnl^2 + 648 \gnl^3  + 384 \fnl^3  \hnl
\right.\right. \right. \notag \\
&
\left.\left. \left.
\qquad + 7776 \fnl \gnl \hnl + 1440 \hnl^2 + 1800 \fnl^2 \inl + 1440 \gnl \inl + 720 \fnl \jnl 
  \right) 
\right.\right. \notag \\
&
\left.\left.
+ R^2 \left( 192 \fnl^6 - 576 \fnl^4 \gnl + 1404  \fnl^2 \gnl^2 -  384 \fnl^3 \hnl + 3456 \fnl \gnl \hnl  
\right.\right. \right. \notag \\
&
\left.\left.\left.
\qquad
+ 720 \hnl^2 + 720 \fnl^2 \inl + 720 \gnl \inl + 360 \fnl \jnl
\right) 
\right]
\left[ \frac{R}{1+R} \calP_\phi(L^{-1})\log(kL) \right]^2
\right\} \, , \notag \\
\end{align}
\begin{align}
\label{eq:gnl-mixed}
\gnl^\text{(tot)} 
= &
\left( \frac{R}{1+R} \right)^3 
\left\{
\gnl
+ \left( \frac35 \right)^2 \left[
 12 \fnl^2 \gnl + 9 \gnl^2 + 24 \fnl \hnl + 10 \inl  
\right.\right. \notag \\
&
\left.\left.
\qquad\qquad\qquad\qquad\qquad
+ R\left( -9 \gnl^2 + 24 \fnl \hnl + 10 \inl \right)
\right]  \frac{R}{1+R}  \calP_\phi(L^{-1})\log(kL) 
\right. 
\notag \\
&
\left.
+\left( \frac35 \right)^4 \left[
216 \fnl^2 \gnl^2 + 135 \gnl^3 + 96 \fnl^3 \hnl + 864 \fnl \gnl \hnl + 144 \hnl^2 + 360 \fnl^2 \inl 
\right.\right. \notag \\
&
\left.\left.
\qquad\qquad
+ 495 \gnl \inl + 360 \fnl \jnl + 105 \knl
\right.\right. \notag \\
&
\left.\left.
+ R \left( 
-144 \fnl^4 \gnl + 108 \fnl^2 \gnl^2 -27 \gnl^3 -96 \fnl^3 \hnl + 1152 \fnl \gnl \hnl +288 \hnl^2 
\right.\right.\right. \notag \\
&
\left.\left.\left.
\qquad\qquad
+ 600 \fnl^2 \inl 
+ 720 \gnl \inl + 720 \fnl \jnl + 210 \knl
\right) 
\right.\right. \notag \\
&
\left.\left.
+ R^2 \left( -48 \fnl^4 \gnl + 180 \fnl^2 \gnl^2 + 54 \gnl^3 
- 192 \fnl^3 \hnl + 288 \fnl \gnl \hnl  + 144 \hnl^2 
\right. \right. \right.
\notag  \\
&
\left.\left.\left.
\qquad\qquad
+  240 \fnl^2 \inl + 225 \gnl \inl + 360 \fnl \jnl + 105 \knl \right)
\right]  \left[  \frac{R}{1+R}  \calP_\phi(L^{-1})\log(kL) \right]^2 
\right\} \, . \notag \\
\end{align}
These non-linear parameters for a given $R$ are shown in Fig.~\ref{fig:2comp_bareNLvsloopNL}.
We can find similar results to those found in the previous section that 
for not too large values of bare non-linear parameters, tree terms dominate.
Meanwhile, as can be seen from Fig.~\ref{fig:2compSYineq}, notice that in this case, 
$\tnl^{\rm (tot)} - (6 \fnl^{\rm (tot)}/5)^2$ significantly deviates from 0 
and the ratio 
$\tnl^{\rm (tot)}/(6 \fnl^{\rm (tot)}/5)^2$ becomes $10^4$ due to a multi-field nature of the model.
Furthermore, around the point at which the bare $\fnl$ is $10^6$, the difference $\tnl^{\rm (tot)} - (6 \fnl^{\rm (tot)}/5)^2$ 
goes negative, which indicates that the Suyama-Yamaguchi (SY) inequality \cite{Suyama:2007bg} breaks down. However, 
this is because the perturbative expansion becomes invalid and the two-loop contribution dominates over 
that from one-loop terms. In other words, it means that the truncation at this order is inappropriate around there.
In Ref.~\cite{Sugiyama:2012tr}, it was shown that if we include all loop contributions, the SY inequality 
holds.

\subsection{Mixed source case II}

Next we consider the case where, unlike (\ref{zetaexpansion-spectator}), the curvature perturbation
associated with the spectator field $\sigma$ has no linear term so that~\cite{nolinear}
\begin{equation}
\label{eq:zeta_mixedII}
\zeta = \zeta_\phi + \frac35 \fnl \zeta_\sigma^2
+\cdots \, .
\end{equation}
The power spectrum in this model is given by 
\begin{align}
P_\zeta = &
P_\phi \left\{ 
1 + 4 R  \left( \frac35 \right)^2 \fnl^2 \left[ R \calP_\phi (L^{-1})\log(kL) \right] \right.
\nonumber\\
& \left. \hspace{0.8cm}
+  \left( \frac35 \right)^4 \left\{ 45 \gnl^2 + 48 \fnl \hnl \right\}  \left[ R \calP_\phi (L^{-1})\log(kL) \right]^2
\right\} \, ,
\end{align}
where we have again defined $R$ as in \eqref{eq:R}. 
However, it should be noted here that $R$ is defined by the ratio of the power spectra 
of $\sigma$ and $\phi$ at the tree level without including the loop terms.  In the case where $\zeta$ is given by 
 Eq.~\eqref{eq:zeta_mixedII}, 
the leading term in the power spectrum of $\sigma$ comes from the loop, thus 
$R$ does not  represent the ratio of the ``actual" power spectra which include the loop terms in this case.
When the power spectrum is dominated by the loop term from $\sigma$, the primordial fluctuations 
become highly non-Gaussian, then it is inconsistent with observations, which we will shortly show below.
Requiring that the loop term be sub-dominant, we require 
\begin{equation}
\left( \frac56 R \fnl \right)^2 \mathcal{P}_\phi \log (kL) \ll 1.
\end{equation}
Thus the combination $R \fnl$ should satisfy
\begin{equation}
\label{eq:RfNL_ungauss}
R \fnl  \ll 10^5.
\end{equation}

Non-linear parameters in this case are given, up to two-loop order, as 
\begin{align}
\fnl^{\rm (tot)} = & 
\frac56 R^2 \left\{
8 \left( \frac35 \right)^3 \fnl^3  \left[ R~ \calP_\phi (L^{-1})\log(kL) \right] 
\right.
\notag \\
&
\left.
+ \left( \frac35 \right)^5 \left( 
240 \fnl^2 \hnl + 450 \fnl \gnl^2 - 64 R \fnl^5 
\right)\left[ R~ \calP_\phi (L^{-1})\log(kL) \right]^2
\right\},   \\
\tnl^{\rm (tot)}
 = &  
 R^3 \left\{
16 \left( \frac35 \right)^4 \fnl^4  \left[ R~ \calP_\phi (L^{-1})\log(kL) \right] 
\right.
\notag \\
&
\left.
+ \left( \frac35 \right)^6 \left( 
3672 \fnl^2 \gnl^2 + 768  \fnl^3 \hnl - 192 R \fnl^6
\right)\left[ R~ \calP_\phi (L^{-1})\log(kL) \right]^2
\right\}, \\
\gnl^{\rm (tot)} 
=&
\frac{R^3 }{6}  \left( \frac35 \right)^4 \left( 
1296 \fnl^2 \gnl^2 + 576  \fnl^3 \hnl 
\right)\left[ R~ \calP_\phi (L^{-1})\log(kL) \right]^2.
\end{align}
When the condition \eqref{eq:RfNL_ungauss} is not satisfied, 
$\fnl^{\rm (tot)} \gg 10^5$, so that the fluctuations are highly non-Gaussian and not allowed by observations.

\begin{figure}[htbp]
  \begin{center}
    \resizebox{160mm}{!}{
     \includegraphics{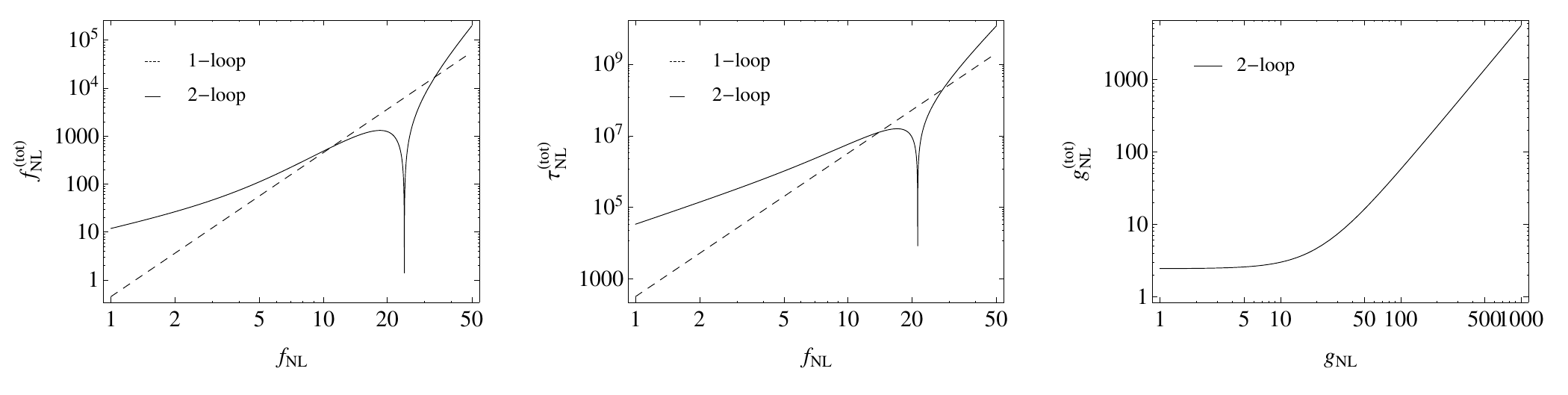}
}
  \end{center}
  \caption{Plot of  $f_{\rm NL}^{\rm (tot)}$ (left), $\tnl^{\rm (tot)}$ (middle) and $\gnl^{\rm (tot)}$ (right)  as a function of 
  bare non-linear parameters.  Here we assume $R=500$. For non-linear parameters, we set 
  $\gnl =10^3$ for the left and middle panels and $\fnl=1$ for the right panel. In all cases, we assume $\hnl=10^3$. 
 }
  \label{fig:ungauss_NL}
\end{figure}

\begin{figure}[htbp]
  \begin{center}
    \resizebox{160mm}{!}{
     \includegraphics{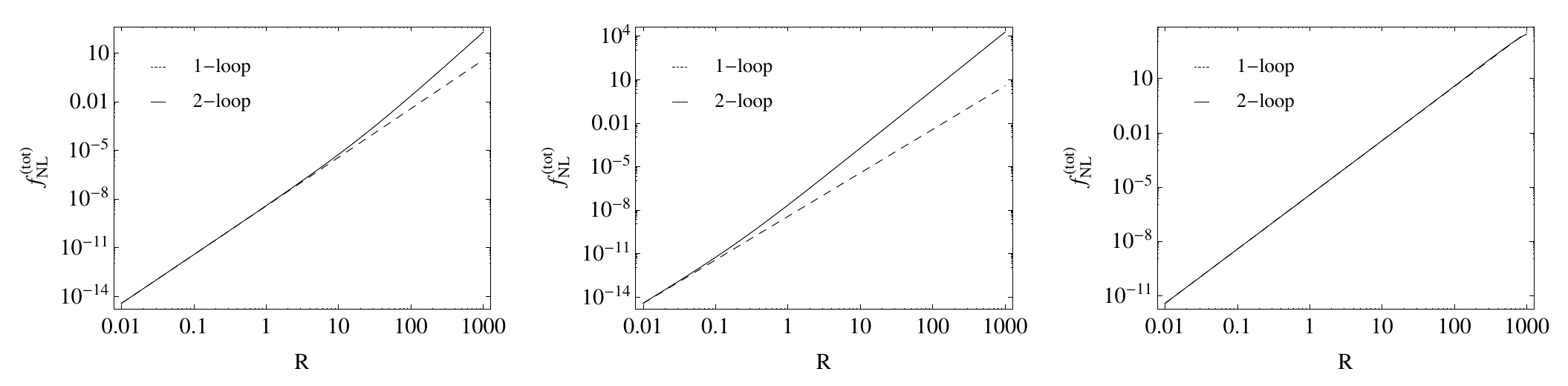}
}
  \end{center}
  \caption{Plot of $\fnl^{\rm (tot)}$ as a function of the ratio $R$ for $\fnl=1$ and $\gnl=10^3$ (left), $\fnl=1$ and $\gnl=10^4$ (middle) and $\fnl=10$ and $\gnl=10^3$ (right). We set $\hnl=10^3$ and other non-linear parameters to vanish for all cases.}
  \label{fig:Rdep}
\end{figure}

\begin{figure}[htbp]
  \begin{center}
    \resizebox{140mm}{!}{
     \includegraphics{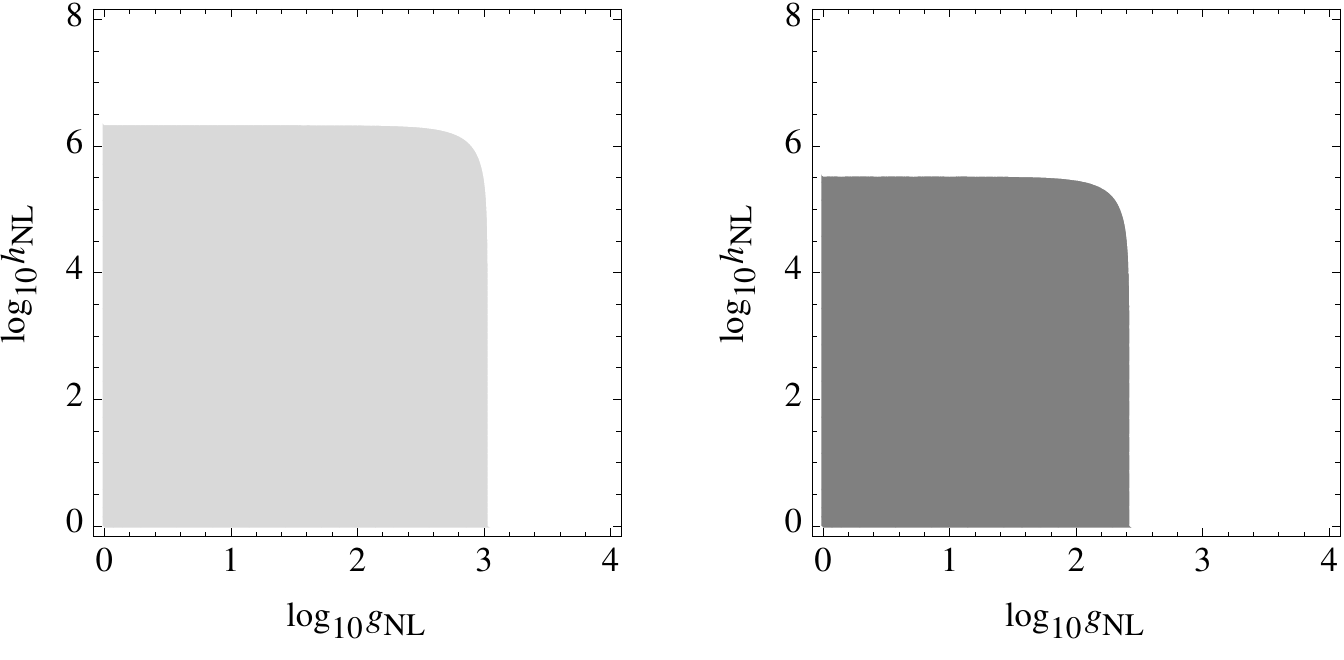}
}
  \end{center}
  \caption{PLANCK bounds for $\fnl^{\rm (tot)}$ (left) and $\tnl^{\rm (tot)}$ (right) for $R=500$ and $\fnl=1$.
  The constraint from $\tnl^{\rm (tot)}$ is stronger than that from $\fnl^{\rm (tot)}$. 
 }
  \label{fig:gnl_hnl_const_ungauss}
\end{figure}

In Fig.~\ref{fig:ungauss_NL},  we show $\fnl^{\rm (tot)}$, $\tnl^{\rm (tot)}$, and $\gnl^{\rm (tot)}$
as a function of bare non-linear parameters. From the plots of $\fnl^{\rm (tot)}$ and 
$\tnl^{\rm (tot)}$, one can notice that, for the case with up to two-loop corrections included, the values of the non-linear parameters
 change their signs from positive to negative at around $\fnl \sim 20$, 
 which indicates that the truncation at the two-loop order may be inappropriate beyond this value of $\fnl$. 
Furthermore, interestingly, two-loop terms can dominate over one-loop terms in the small $\fnl$ region, as seen from Fig.~\ref{fig:ungauss_NL}
when we assume relatively large value of $R$. 
In Fig.~\ref{fig:Rdep}, we show the dependence of $\fnl^\text{(tot)}$ on $R$. The two-loop terms contribute to $\fnl^\text{(tot)}$ appreciably for small values of $\fnl$ when $R$ is relatively large. 
This contribution is coming mainly
from $\fnl\gnl^2$ terms, which can be larger than $\fnl^2\hnl$ 
if $\fnl$ is small and $\gnl\sim\hnl$.
Notice that the value for $R$ assumed in the figure satisfies the
 relation \eqref{eq:RfNL_ungauss}: thus, the power spectrum is still dominated 
by that from $\phi$, which is assumed to be Gaussian here. 
Since this effect comes from the higher order non-linear parameters such as $\gnl$ and $\hnl$, we can constrain these parameters 
from the PLANCK constraints on $\fnl^{\rm (tot)}$ and $\tnl^{\rm (tot)}$, which is shown in Fig.~\ref{fig:gnl_hnl_const_ungauss}, assuming $R=500$ 
and $\fnl=1$. In the figure, shaded regions indicate 
those satisfying $ - 8.9 < \fnl^{\rm (tot)} < 14.3$ and $\tnl^{\rm (tot)} < 2800$, which correspond to 95~\% constraints from PLANCK.
The figure  illustrates  that, in this kind of model, higher order non-linear parameters can 
be strongly constrained from those for lower order 
counterparts. 
However, we note here that we have included up to the two-loop order terms in this article, and 
if we include higher order loops such as three loop, the results might be affected. 
This may be interesting to 
investigate, but it is beyond the scope of this article and is left for the future work.

\section{Conclusion and Discussion}
\label{sec:conclusion}

We have investigated how higher order non-linear parameters affect the lower order ones
through the loop corrections. We have explicitly calculated the corrections for $\fnl^{\rm (tot)}$, $\tnl^{\rm (tot)}$, and $\gnl^{\rm (tot)}$ 
up to two-loop order for single-source and multi-source cases, which have been discussed in the Sections~\ref{sec:formalism} and
\ref{sec:probe}, respectively.

First of all, by explicitly calculating the loop contributions up to two-loop order, we have argued that 
as long as the bare non-linear parameters are not too large to harm the perturbative expansion
of the curvature perturbation, the loop corrections remain very small and the tree contributions
are dominant in determining the total observable ones. This is because of the smallness of the
loop factor, $\calP_\zeta(L^{-1})\log(kL) \sim 2.5 \times 10^{-9}$. One may increase 
the fictitious box size $L$ to have a large logarithmic factor, but then we can no longer resort to
the value of the power spectrum $\calP_\zeta$ constrained in the observable patch.
However, an interesting possibility to avoid this is when the curvature perturbation is sourced by
another field, where we have an additional factor $R$. In such a case, the two-loop terms can
be important as shown in Fig.~\ref{fig:ungauss_NL}.

Furthermore, by looking at the expressions of non-linear parameters including loop corrections, 
we can easily see that, in principle, higher order 
non-linear parameters can be constrained by the observations of lower order counterparts. 
Although in general, such a constraint is weak, recent PLANCK results can give a constraint on 
the bare value of $\gnl$  from the constraint on $\tnl$. 
Interestingly, the bound for the bare $\gnl$ derived from the PLANCK $\tnl$ constraint $(\tnl < 2800$ at 95~\% C.L.) is 
$\gnl \le 5 \times 10^5$, which is almost the same as what is directly obtained from trispectrum observations such as WMAP.

In this paper, we have focused on the case of the local type. We note that higher loop corrections 
can also affect the observables in the non-local form cases.

Now, PLANCK data put a stringent constraint on $\fnl$, however, a non-Gaussian signature of primordial 
density perturbation may come from a higher order one (see Ref.~\cite{Suyama:2013nva} for such an example).
In such a case, the results presented in this article should be useful to investigate such a scenario.
In addition, when severer constraints on lower order non-linear parameters are obtained in future cosmological observations,
we may be able to have more stringent bounds  for higher order non-linear parameters from such lower order constraints.

\section*{Acknowledgments}

We thank Christian Byrnes and Sami Nurmi for useful discussions.
TT would like to thank APCTP for the hospitality
during the visit, where this work was initiated.
JG acknowledges the Max-Planck-Gesellschaft,
the Korea Ministry of Education, Science and Technology, 
Gyeongsangbuk-Do and Pohang City for
the support of the Independent Junior Research Group at the Asia Pacific Center for Theoretical
Physics.
JG is also supported by a Starting Grant through the Basic Science Research Program of 
the National Research Foundation of Korea (2013R1A1A1006701).
The work of TT is partially supported by a Grant-in-Aid for Scientific
Research from the Ministry of Education, Science, Sports, and
Culture, Japan, No.~23740195.

\pagebreak
\appendix 
\noindent
{\bf \Large Appendix}
\allowdisplaybreaks

\section{Expressions for the power spectrum}
\label{app:power}

The one-loop terms for the power spectrum are
\begin{align}
P_\zeta^{(13)}(k) = & 6 \left( \frac{3}{5} \right)^2  \gnl P_\zeta(k) \left[ \calP_\zeta(L^{-1}) \log(kL) \right]\, ,
\\
P_\zeta^{(22)}(k) = & 2 \left( \frac{3}{5} \right)^2  \fnl^2 P_\zeta(k) \left[ \calP_\zeta(L^{-1}) \log(kL) \right]\, .
\end{align}
Meanwhile, the two-loop terms are given by
\begin{align}
P_\zeta^{(15)}(k) = & 30 \left( \frac{3}{5} \right)^4 \inl P_\zeta(k) \left[ \calP_\zeta(L^{-1}) \log(kL) \right]^2 \, ,
\\
P_\zeta^{(24)}(k) = & 48 \left( \frac{3}{5} \right)^4 \fnl\hnl P_\zeta(k) \left[ \calP_\zeta(L^{-1}) \log(kL) \right]^2 \, ,
\\
P_\zeta^{(33)}(k) = & 45 \left( \frac{3}{5} \right)^4 \gnl^2 P_\zeta(k) \left[ \calP_\zeta(L^{-1}) \log(kL) \right]^2 \, .
\end{align}

\section{Expressions for the bispectrum}
\label{app:bi}

\begin{figure}[htbp]
  \begin{center}
    \rotatebox{0}{\resizebox{180mm}{!}{
     \includegraphics{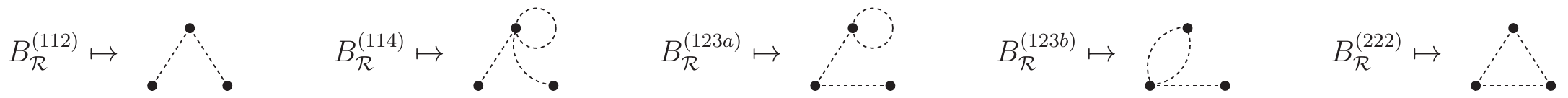}}
}
  \end{center}
  \caption{Connected structure of the bispectrum up to one-loop corrections.}
  \label{fig:bispect}
\end{figure}

\begin{figure}[htbp]
  \begin{center}
    \rotatebox{0}{\resizebox{120mm}{!}{
     \includegraphics{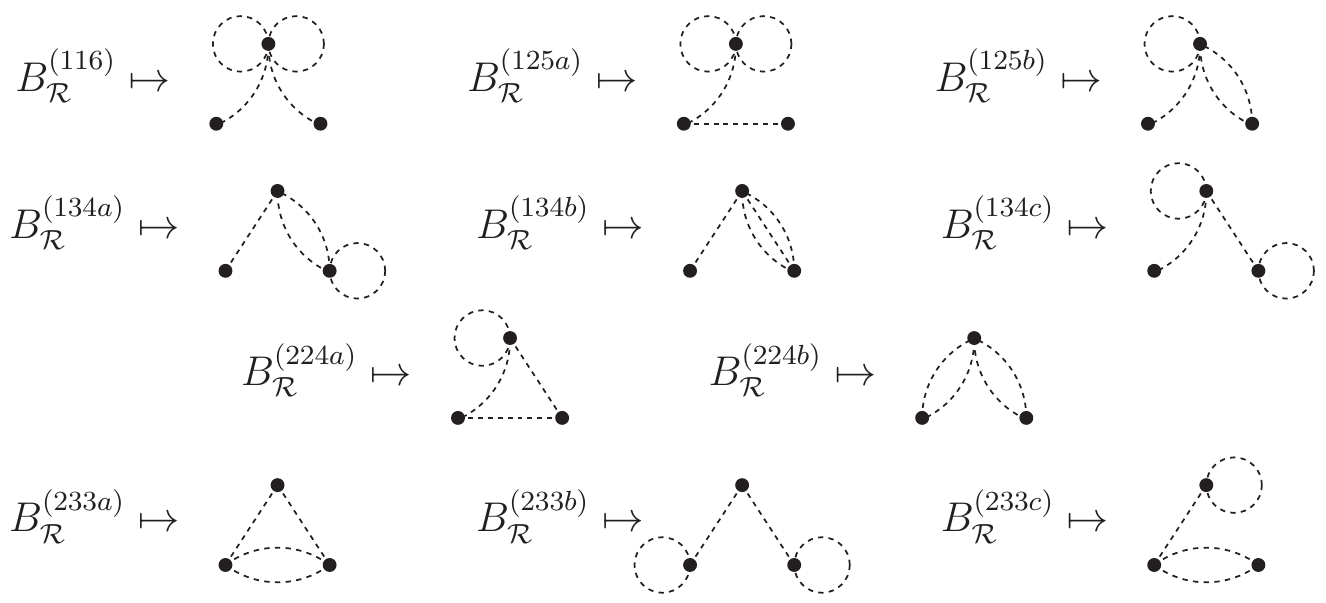}}
}
  \end{center}
  \caption{Connected structure of the bispectrum at two-loop order.}
  \label{fig:bispect2}
\end{figure}

Up to the one-loop corrections to the bispectrum we have
\begin{align}
B_\zeta^{(112)} & = \frac{6}{5}\fnl \left[ P_\zeta(k_1)P_\zeta(k_2) + \text{2 perm} \right] \, ,
\\
B_\zeta^{(222)} & = 8\left(\frac{3}{5} \right)^3\fnl^3 \left[ P_\zeta(k_1)P_\zeta(k_2) + \text{2 perm} \right] \calP_\zeta(L^{-1})\log(kL) \, ,
\\
B_\zeta^{(123a)} & = 12 \left( \frac{3}{5} \right)^3 \fnl\gnl \left[ P_\zeta(k_1)P_\zeta(k_2) + \text{2 perm} \right] \calP_\zeta(L^{-1})\log(kL) \, ,
\\
B_\zeta^{(123b)} & = 24 \left( \frac{3}{5} \right)^3 \fnl\gnl \left[ P_\zeta(k_1)P_\zeta(k_2) + \text{2 perm} \right] \calP_\zeta(L^{-1})\log(kL) \, ,
\\
B_\zeta^{(114)} & = 12 \left( \frac{3}{5} \right)^3 \hnl \left[ P_\zeta(k_1)P_\zeta(k_2) + \text{2 perm} \right] \calP_\zeta(L^{-1})\log(kL) \, .
\end{align}
The connected structure of these terms is shown in Fig.~\ref{fig:bispect}. Likewise, at the two-loop order the contributions are
\begin{align}
B^{(116)}_\zeta = & 90 \left( \frac{3}{5} \right)^5 \jnl \left[ P_\zeta(k_1)P_\zeta(k_2) + \text{2 perm} \right] \left[ \calP_\zeta(L^{-1})\log(kL) \right]^2 \, ,
\\
B^{(125a)}_\zeta = & 60 \left( \frac{3}{5} \right)^5 \fnl\inl \left[ P_\zeta(k_1)P_\zeta(k_2) + \text{2 perm} \right] \left[ \calP_\zeta(L^{-1})\log(kL) \right]^2 \, ,
\\
B^{(125b)}_\zeta = & 240 \left( \frac{3}{5} \right)^5 \fnl\inl \left[ P_\zeta(k_1)P_\zeta(k_2) + \text{2 perm} \right] \left[ \calP_\zeta(L^{-1})\log(kL) \right]^2 \, ,
\\
B^{(134a)}_\zeta = & 144 \left( \frac{3}{5} \right)^5 \gnl\hnl \left[ P_\zeta(k_1)P_\zeta(k_2) + \text{2 perm} \right] \left[ \calP_\zeta(L^{-1})\log(kL) \right]^2 \, ,
\\
B^{(134b)}_\zeta = & 144 \left( \frac{3}{5} \right)^5 \gnl\hnl \left[ P_\zeta(k_1)P_\zeta(k_2) + \text{2 perm} \right] \left[ \calP_\zeta(L^{-1})\log(kL) \right]^2 \, ,
\\
B^{(134c)}_\zeta = & 72 \left( \frac{3}{5} \right)^5 \gnl\hnl \left[ P_\zeta(k_1)P_\zeta(k_2) + \text{2 perm} \right] \left[ \calP_\zeta(L^{-1})\log(kL) \right]^2 \, ,
\\
B^{(224a)}_\zeta = & 144 \left( \frac{3}{5} \right)^5 \fnl^2\hnl \left[ P_\zeta(k_1)P_\zeta(k_2) + \text{2 perm} \right] \left[ \calP_\zeta(L^{-1})\log(kL) \right]^2 \, ,
\\
B^{(224b)}_\zeta = & 96 \left( \frac{3}{5} \right)^5 \fnl^2\hnl \left[ P_\zeta(k_1)P_\zeta(k_2) + \text{2 perm} \right] \left[ \calP_\zeta(L^{-1})\log(kL) \right]^2 \, ,
\\
B^{(332a)}_\zeta = & 360 \left( \frac{3}{5} \right)^5 \fnl\gnl^2 \left[ P_\zeta(k_1)P_\zeta(k_2) + \text{2 perm} \right] \left[ \calP_\zeta(L^{-1})\log(kL) \right]^2 \, ,
\\
B^{(332b)}_\zeta = & 18 \left( \frac{3}{5} \right)^5 \fnl\gnl^2 \left[ P_\zeta(k_1)P_\zeta(k_2) + \text{2 perm} \right] \left[ \calP_\zeta(L^{-1})\log(kL) \right]^2 \, ,
\\
B^{(332c)}_\zeta = & 72 \left( \frac{3}{5} \right)^5 \fnl\gnl^2 \left[ P_\zeta(k_1)P_\zeta(k_2) + \text{2 perm} \right] \left[ \calP_\zeta(L^{-1})\log(kL) \right]^2 \, ,
\end{align}
and the connected structure is as shown in Fig.~\ref{fig:bispect2}.

\section{Expressions for the trispectrum}
\label{app:tri}

\begin{figure}[htbp]
  \begin{center}
    \rotatebox{0}{\resizebox{180mm}{!}{
     \includegraphics{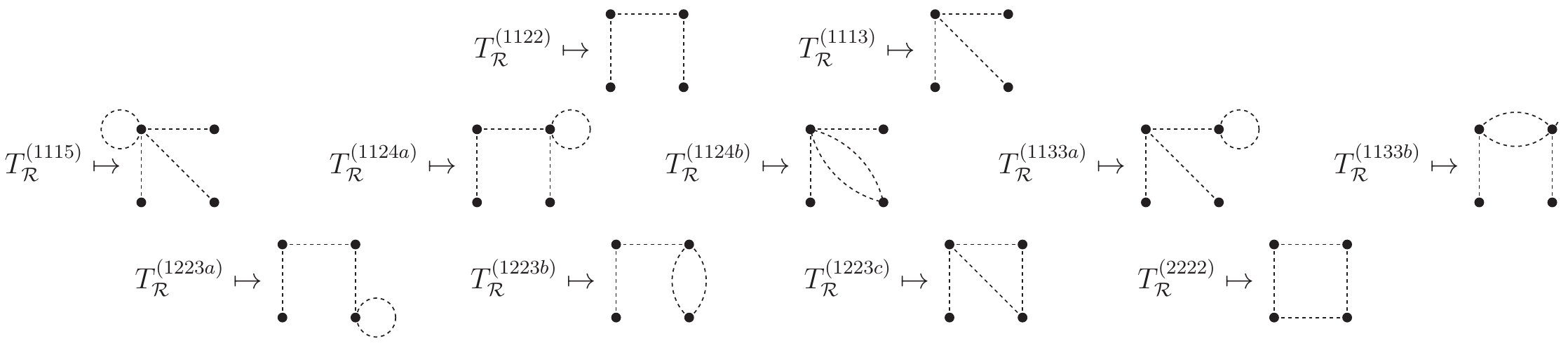}}
}
  \end{center}
  \caption{Connected structure of the trispectrum up to one loop.}
  \label{fig:trispect}
\end{figure}

The two contributions to the tree trispectrum are
\begin{align}
T_\zeta^{(1122)} & = 2 \left( \frac{3}{5} \right)^2 \fnl^2 \left\{ P_\zeta(k_1)P_\zeta(k_2) \left[ P_\zeta(k_{13}) + P_\zeta(k_{14}) \right] + \text{11 perm} \right\} \, ,
\\
T_\zeta^{(1113)} & = \frac{54}{25}\gnl \left[ P_\zeta(k_1)P_\zeta(k_2)P_\zeta(k_3) + \text{3 perm} \right] \, ,
\end{align}
and the one-loop corrections are given by
\begin{align}
T_\zeta^{(2222)} & = 8 \left( \frac{3}{5} \right)^4 \fnl^4 \left\{ P_\zeta(k_1)P_\zeta(k_2) \left[ P_\zeta(k_{13}) + P_\zeta(k_{14}) \right] + \text{11 perm} \right\} 
\nonumber\\
& \hspace{7em} \times \calP_\zeta(L^{-1})\log(kL) \, ,
\\
T_\zeta^{(1223a)} & = 12 \left( \frac{3}{5} \right)^4 \fnl^2\gnl \left\{ P_\zeta(k_1)P_\zeta(k_2) \left[ P_\zeta(k_{13}) + P_\zeta(k_{14}) \right] + \text{11 perm} \right\} 
\nonumber\\
& \hspace{9em} \times \calP_\zeta(L^{-1})\log(kL) \, ,
\\
T_\zeta^{(1223b)} & = 24 \left( \frac{3}{5} \right)^4 \fnl^2\gnl \left\{ P_\zeta(k_1)P_\zeta(k_2) \left[ P_\zeta(k_{13}) + P_\zeta(k_{14}) \right] + \text{11 perm} \right\} 
\nonumber\\
& \hspace{9em} \times \calP_\zeta(L^{-1})\log(kL) \, ,
\\
T_\zeta^{(1223c)} & = 24 \left( \frac{3}{5} \right)^4 \fnl^2\gnl \Big( P_\zeta(k_1) \big\{ P_\zeta(k_2)P_\zeta(k_3) + \left[ P_\zeta(k_2) + P_\zeta(k_3) \right] P_\zeta(k_{23}) \big\} 
\nonumber\\
& \hspace{9em} + \text{11 perm} \Big) \calP_\zeta(L^{-1})\log(kL) \, ,
\\
T_\zeta^{(1133a)} & = 54 \left( \frac{3}{5} \right)^4 \gnl^2 \left[ P_\zeta(k_1)P_\zeta(k_2)P_\zeta(k_3) + \text{3 perm} \right] \calP_\zeta(L^{-1})\log(kL) \, ,
\\
T_\zeta^{(1133b)} & = 18 \left( \frac{3}{5} \right)^4 \gnl^2 \left\{ P_\zeta(k_1)P_\zeta(k_2) \left[ P_\zeta(k_{13}) + P_\zeta(k_{14}) \right] + \text{11 perm} \right\} 
\nonumber\\
& \hspace{7em} \times \calP_\zeta(L^{-1})\log(kL) \, ,
\\
T_\zeta^{(1124a)} & = 24 \left( \frac{3}{5} \right)^4 \fnl\hnl \left\{ P_\zeta(k_1)P_\zeta(k_2) \left[ P_\zeta(k_{13}) + P_\zeta(k_{14}) \right] + \text{11 perm} \right\} 
\nonumber\\
& \hspace{9em} \times \calP_\zeta(L^{-1})\log(kL) \, ,
\\
T_\zeta^{(1124b)} & = 144 \left( \frac{3}{5} \right)^4 \fnl\hnl \left[ P_\zeta(k_1)P_\zeta(k_2)P_\zeta(k_3) + \text{3 perm} \right] \calP_\zeta(L^{-1})\log(kL) \, ,
\\
T_\zeta^{(1115)} & = 60 \left( \frac{3}{5} \right)^4 i_{\rm NL} \left[ P_\zeta(k_1)P_\zeta(k_2)P_\zeta(k_3) + \text{3 perm} \right] \calP_\zeta(L^{-1})\log(kL) \, ,
\end{align}
The connected structure of these terms is shown in Fig.~\ref{fig:trispect}.

\begin{figure}[htbp]
  \begin{center}
    \rotatebox{0}{\resizebox{180mm}{!}{
     \includegraphics{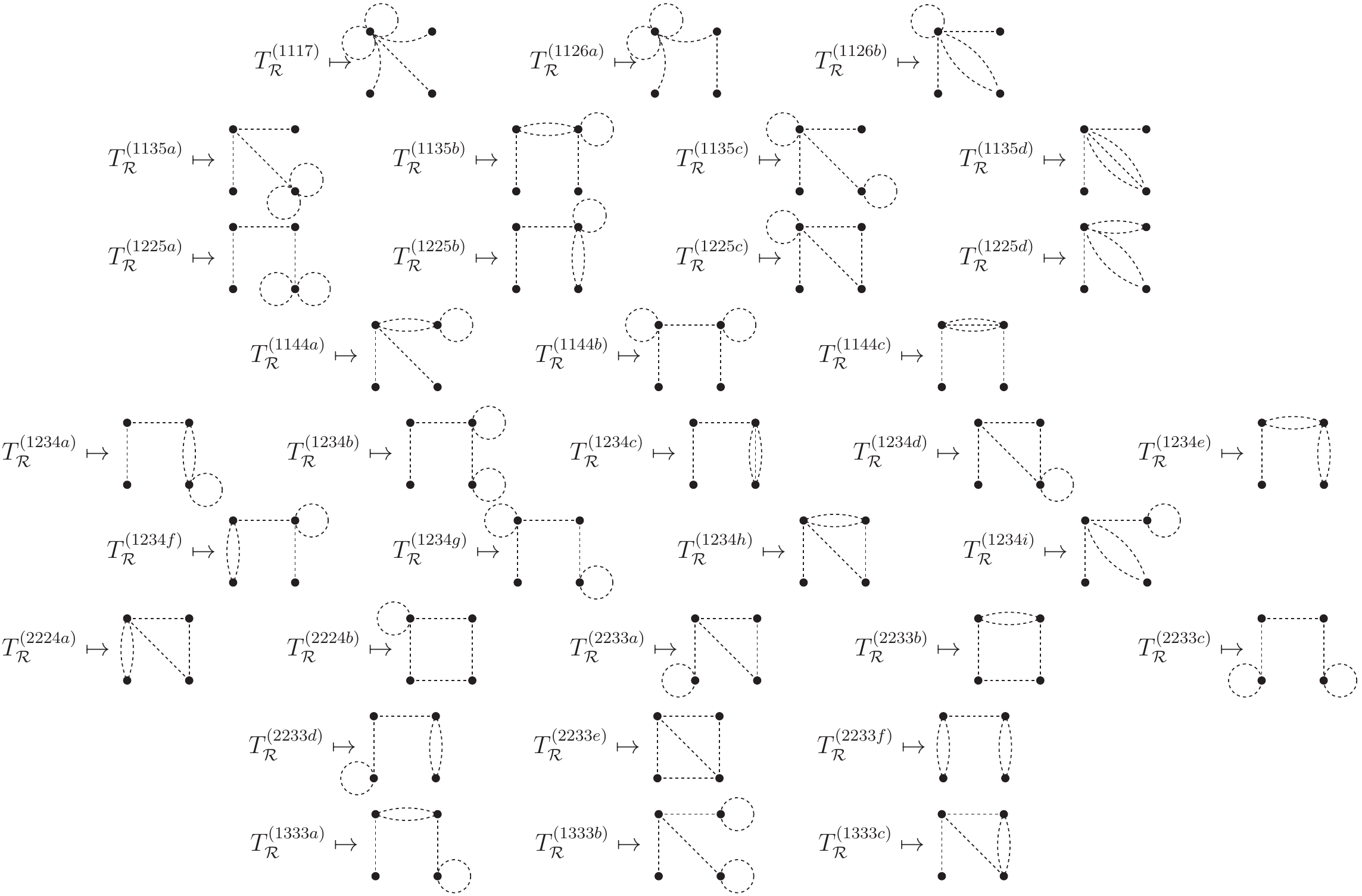}}
}
  \end{center}
  \caption{Connected structure of the trispectrum at two-loop order.}
  \label{fig:trispect2}
\end{figure}

\bigskip
The terms contributing at the two-loop level are the following:
\begin{align}
T^{(1117)}_\zeta = & 630 \left( \frac{3}{5} \right)^6 \knl \left[ P_\zeta(k_1)P_\zeta(k_2)P_\zeta(k_3) + \text{3 perm} \right] \left[ \calP_\zeta(L^{-1})\log(kL) \right]^2 \, ,
\\
T_\zeta^{(1126a)} = & 180 \left( \frac{3}{5} \right)^6 \fnl\jnl \left\{ P_\zeta(k_1)P_\zeta(k_2) \left[ P_\zeta(k_{13}) + P_\zeta(k_{14}) \right] + \text{11 perm} \right\} \left[ \calP_\zeta(L^{-1})\log(kL) \right]^2 \, ,
\\
T^{(1126b)}_\zeta = & 2160 \left( \frac{3}{5} \right)^6 \fnl\jnl \left[ P_\zeta(k_1)P_\zeta(k_2)P_\zeta(k_3) + \text{3 perm} \right] \left[ \calP_\zeta(L^{-1})\log(kL) \right]^2 \, ,
\\
T^{(1135a)}_\zeta = & 270 \left( \frac{3}{5} \right)^6 \gnl\inl \left[ P_\zeta(k_1)P_\zeta(k_2)P_\zeta(k_3) + \text{3 perm} \right] \left[ \calP_\zeta(L^{-1})\log(kL) \right]^2 \, ,
\\
T_\zeta^{(1135b)} = & 360 \left( \frac{3}{5} \right)^6 \gnl\inl \left\{ P_\zeta(k_1)P_\zeta(k_2) \left[ P_\zeta(k_{13}) + P_\zeta(k_{14}) \right] + \text{11 perm} \right\} \left[ \calP_\zeta(L^{-1})\log(kL) \right]^2 \, ,
\\
T^{(1135c)}_\zeta = & 540 \left( \frac{3}{5} \right)^6 \gnl\inl \left[ P_\zeta(k_1)P_\zeta(k_2)P_\zeta(k_3) + \text{3 perm} \right] \left[ \calP_\zeta(L^{-1})\log(kL) \right]^2 \, ,
\\
T^{(1135d)}_\zeta = & 2160 \left( \frac{3}{5} \right)^6 \fnl\jnl \left[ P_\zeta(k_1)P_\zeta(k_2)P_\zeta(k_3) + \text{3 perm} \right] \left[ \calP_\zeta(L^{-1})\log(kL) \right]^2 \, ,
\\
T_\zeta^{(1225a)} = & 60 \left( \frac{3}{5} \right)^6 \fnl^2\inl \left\{ P_\zeta(k_1)P_\zeta(k_2) \left[ P_\zeta(k_{13}) + P_\zeta(k_{14}) \right] + \text{11 perm} \right\} \left[ \calP_\zeta(L^{-1})\log(kL) \right]^2 \, ,
\\
T_\zeta^{(1225b)} = & 240 \left( \frac{3}{5} \right)^6 \fnl^2\inl \left\{ P_\zeta(k_1)P_\zeta(k_2) \left[ P_\zeta(k_{13}) + P_\zeta(k_{14}) \right] + \text{11 perm} \right\} \left[ \calP_\zeta(L^{-1})\log(kL) \right]^2 \, ,
\\
T_\zeta^{(1225c)} = & 720 \left( \frac{3}{5} \right)^6 \fnl^2\inl \left[ P_\zeta(k_1)P_\zeta(k_2)P_\zeta(k_3) + \text{3 perm} \right] \left[ \calP_\zeta(L^{-1})\log(kL) \right]^2
\nonumber\\
& + 240 \left( \frac{3}{5} \right)^6 \fnl^2\inl \left\{ P_\zeta(k_1)P_\zeta(k_2) \left[ P_\zeta(k_{13}) + P_\zeta(k_{14}) \right] + \text{11 perm} \right\} \left[ \calP_\zeta(L^{-1})\log(kL) \right]^2 \, ,
\\
T_\zeta^{(1225d)} = & 1440 \left( \frac{3}{5} \right)^6 \fnl^2\inl \left[ P_\zeta(k_1)P_\zeta(k_2)P_\zeta(k_3) + \text{3 perm} \right] \left[ \calP_\zeta(L^{-1})\log(kL) \right]^2
\\
T_\zeta^{(1144a)} = & 864 \left( \frac{3}{5} \right)^6 \hnl^2 \left[ P_\zeta(k_1)P_\zeta(k_2)P_\zeta(k_3) + \text{3 perm} \right] \left[ \calP_\zeta(L^{-1})\log(kL) \right]^2
\\
T_\zeta^{(1144b)} = & 72 \left( \frac{3}{5} \right)^6 \hnl^2 \left\{ P_\zeta(k_1)P_\zeta(k_2) \left[ P_\zeta(k_{13}) + P_\zeta(k_{14}) \right] + \text{11 perm} \right\} \left[ \calP_\zeta(L^{-1})\log(kL) \right]^2 \, ,
\\
T_\zeta^{(1144c)} = & 288 \left( \frac{3}{5} \right)^6 \hnl^2 \left\{ P_\zeta(k_1)P_\zeta(k_2) \left[ P_\zeta(k_{13}) + P_\zeta(k_{14}) \right] + \text{11 perm} \right\} \left[ \calP_\zeta(L^{-1})\log(kL) \right]^2 \, ,
\\
T_\zeta^{(1234a)} = & 144 \left( \frac{3}{5} \right)^6 \fnl\gnl\hnl \left\{ P_\zeta(k_1)P_\zeta(k_2) \left[ P_\zeta(k_{13}) + P_\zeta(k_{14}) \right] + \text{11 perm} \right\} \left[ \calP_\zeta(L^{-1})\log(kL) \right]^2 \, ,
\\
T_\zeta^{(1234b)} = & 72 \left( \frac{3}{5} \right)^6 \fnl\gnl\hnl \left\{ P_\zeta(k_1)P_\zeta(k_2) \left[ P_\zeta(k_{13}) + P_\zeta(k_{14}) \right] + \text{11 perm} \right\} \left[ \calP_\zeta(L^{-1})\log(kL) \right]^2 \, ,
\\
T_\zeta^{(1234c)} = & 288 \left( \frac{3}{5} \right)^6 \fnl\gnl\hnl \left\{ P_\zeta(k_1)P_\zeta(k_2) \left[ P_\zeta(k_{13}) + P_\zeta(k_{14}) \right] + \text{11 perm} \right\} \left[ \calP_\zeta(L^{-1})\log(kL) \right]^2 \, ,
\\
T_\zeta^{(1234d)} = & 864 \left( \frac{3}{5} \right)^6 \fnl\gnl\hnl \left[ P_\zeta(k_1)P_\zeta(k_2)P_\zeta(k_3) + \text{3 perm} \right] \left[ \calP_\zeta(L^{-1})\log(kL) \right]^2
\nonumber\\
& + 288 \left( \frac{3}{5} \right)^6 \fnl\gnl\hnl \left\{ P_\zeta(k_1)P_\zeta(k_2) \left[ P_\zeta(k_{13}) + P_\zeta(k_{14}) \right] + \text{11 perm} \right\} \left[ \calP_\zeta(L^{-1})\log(kL) \right]^2 \, ,
\\
T_\zeta^{(1234f)} = & 288 \left( \frac{3}{5} \right)^6 \fnl\gnl\hnl \left\{ P_\zeta(k_1)P_\zeta(k_2) \left[ P_\zeta(k_{13}) + P_\zeta(k_{14}) \right] + \text{11 perm} \right\} \left[ \calP_\zeta(L^{-1})\log(kL) \right]^2 \, ,
\\
T_\zeta^{(1234g)} = & 144 \left( \frac{3}{5} \right)^6 \fnl\gnl\hnl \left\{ P_\zeta(k_1)P_\zeta(k_2) \left[ P_\zeta(k_{13}) + P_\zeta(k_{14}) \right] + \text{11 perm} \right\} \left[ \calP_\zeta(L^{-1})\log(kL) \right]^2 \, ,
\\
T_\zeta^{(1234h)} = & 72 \left( \frac{3}{5} \right)^6 \fnl\gnl\hnl \left\{ P_\zeta(k_1)P_\zeta(k_2) \left[ P_\zeta(k_{13}) + P_\zeta(k_{14}) \right] + \text{11 perm} \right\} \left[ \calP_\zeta(L^{-1})\log(kL) \right]^2 \, ,
\\
T_\zeta^{(1234i)} = & 3456 \left( \frac{3}{5} \right)^6 \fnl\gnl\hnl \left[ P_\zeta(k_1)P_\zeta(k_2)P_\zeta(k_3) + \text{3 perm} \right] \left[ \calP_\zeta(L^{-1})\log(kL) \right]^2
\nonumber\\
& + 864 \left( \frac{3}{5} \right)^6 \fnl\gnl\hnl \left\{ P_\zeta(k_1)P_\zeta(k_2) \left[ P_\zeta(k_{13}) + P_\zeta(k_{14}) \right] + \text{11 perm} \right\} \left[ \calP_\zeta(L^{-1})\log(kL) \right]^2 \, ,
\\
T_\zeta^{(1234l)} = & 864 \left( \frac{3}{5} \right)^6 \fnl\gnl\hnl \left[ P_\zeta(k_1)P_\zeta(k_2)P_\zeta(k_3) + \text{3 perm} \right] \left[ \calP_\zeta(L^{-1})\log(kL) \right]^2 \, ,
\\
T_\zeta^{(4222a)} = & 576 \left( \frac{3}{5} \right)^6 \fnl^3\hnl \left[ P_\zeta(k_1)P_\zeta(k_2)P_\zeta(k_3) + \text{3 perm} \right] \left[ \calP_\zeta(L^{-1})\log(kL) \right]^2
\nonumber\\
& + 192 \left( \frac{3}{5} \right)^6 \fnl^3\hnl \left\{ P_\zeta(k_1)P_\zeta(k_2) \left[ P_\zeta(k_{13}) + P_\zeta(k_{14}) \right] + \text{11 perm} \right\} \left[ \calP_\zeta(L^{-1})\log(kL) \right]^2 \, ,
\\
T_\zeta^{(4222b)} = & 192 \left( \frac{3}{5} \right)^6 \fnl^3\hnl \left\{ P_\zeta(k_1)P_\zeta(k_2) \left[ P_\zeta(k_{13}) + P_\zeta(k_{14}) \right] + \text{11 perm} \right\} \left[ \calP_\zeta(L^{-1})\log(kL) \right]^2 \, ,
\\
T_\zeta^{(2233a)} = & 432 \left( \frac{3}{5} \right)^6 \fnl^2\gnl^2 \left[ P_\zeta(k_1)P_\zeta(k_2)P_\zeta(k_3) + \text{3 perm} \right] \left[ \calP_\zeta(L^{-1})\log(kL) \right]^2
\nonumber\\
& + 144 \left( \frac{3}{5} \right)^6 \fnl^2\gnl^2 \left\{ P_\zeta(k_1)P_\zeta(k_2) \left[ P_\zeta(k_{13}) + P_\zeta(k_{14}) \right] + \text{11 perm} \right\} \left[ \calP_\zeta(L^{-1})\log(kL) \right]^2 \, ,
\\
T_\zeta^{(2233b)} = & 1008 \left( \frac{3}{5} \right)^6 \fnl^2\gnl^2 \left\{ P_\zeta(k_1)P_\zeta(k_2) \left[ P_\zeta(k_{13}) + P_\zeta(k_{14}) \right] + \text{11 perm} \right\} \left[ \calP_\zeta(L^{-1})\log(kL) \right]^2 \, ,
\\
T_\zeta^{(2233c)} = & 36 \left( \frac{3}{5} \right)^6 \fnl^2\gnl^2 \left\{ P_\zeta(k_1)P_\zeta(k_2) \left[ P_\zeta(k_{13}) + P_\zeta(k_{14}) \right] + \text{11 perm} \right\} \left[ \calP_\zeta(L^{-1})\log(kL) \right]^2 \, ,
\\
T_\zeta^{(2233d)} = & 72 \left( \frac{3}{5} \right)^6 \fnl^2\gnl^2 \left\{ P_\zeta(k_1)P_\zeta(k_2) \left[ P_\zeta(k_{13}) + P_\zeta(k_{14}) \right] + \text{11 perm} \right\} \left[ \calP_\zeta(L^{-1})\log(kL) \right]^2 \, ,
\\
T_\zeta^{(2233e)} = & 864 \left( \frac{3}{5} \right)^6 \fnl^2\gnl^2 \left[ P_\zeta(k_1)P_\zeta(k_2)P_\zeta(k_3) + \text{3 perm} \right] \left[ \calP_\zeta(L^{-1})\log(kL) \right]^2
\nonumber\\
& + 432 \left( \frac{3}{5} \right)^6 \fnl^2\gnl^2 \left\{ P_\zeta(k_1)P_\zeta(k_2) \left[ P_\zeta(k_{13}) + P_\zeta(k_{14}) \right] + \text{11 perm} \right\} \left[ \calP_\zeta(L^{-1})\log(kL) \right]^2 \, ,
\\
T_\zeta^{(2233g)} = & 144 \left( \frac{3}{5} \right)^6 \fnl^2\gnl^2 \left\{ P_\zeta(k_1)P_\zeta(k_2) \left[ P_\zeta(k_{13}) + P_\zeta(k_{14}) \right] + \text{11 perm} \right\} \left[ \calP_\zeta(L^{-1})\log(kL) \right]^2 \, ,
\\
T_\zeta^{(1333a)} = & 108 \left( \frac{3}{5} \right)^6 \gnl^3 \left\{ P_\zeta(k_1)P_\zeta(k_2) \left[ P_\zeta(k_{13}) + P_\zeta(k_{14}) \right] + \text{11 perm} \right\} \left[ \calP_\zeta(L^{-1})\log(kL) \right]^2 \, ,
\\
T_\zeta^{(1333b)} = & 162 \left( \frac{3}{5} \right)^6 \gnl^3 \left[ P_\zeta(k_1)P_\zeta(k_2)P_\zeta(k_3) + \text{3 perm} \right] \left[ \calP_\zeta(L^{-1})\log(kL) \right]^2 \, ,
\\
T_\zeta^{(1333c)} = & 648 \left( \frac{3}{5} \right)^6 \gnl^3 \left[ P_\zeta(k_1)P_\zeta(k_2)P_\zeta(k_3) + \text{3 perm} \right] \left[ \calP_\zeta(L^{-1})\log(kL) \right]^2
\nonumber\\
& + 216 \left( \frac{3}{5} \right)^6 \gnl^3 \left\{ P_\zeta(k_1)P_\zeta(k_2) \left[ P_\zeta(k_{13}) + P_\zeta(k_{14}) \right] + \text{11 perm} \right\} \left[ \calP_\zeta(L^{-1})\log(kL) \right]^2 \, .
\end{align}
For the connected structure, see Fig.~\ref{fig:trispect2}.

\end{document}